# An Interdisciplinary Survey on Information Flows in Supply Chains

JAN PENNEKAMP and ROMAN MATZUTT, RWTH Aachen University, Germany
CHRISTOPHER KLINKMÜLLER, BPMotion, Australia
LENNART BADER and MARTIN SERROR, Fraunhofer FKIE, Germany
ERIC WAGNER, Fraunhofer FKIE, Germany and RWTH Aachen University, Germany
SIDRA MALIK, Data61 CSIRO, Australia
MARIA SPIß and JESSICA RAHN, Institute for Industrial Management at RWTH Aachen University, Germany
TAN GÜRPINAR, Fraunhofer IML, Germany
EDUARD VLAD and SANDER J. J. LEEMANS, RWTH Aachen University, Germany
SALIL S. KANHERE, University of New South Wales, Australia
VOLKER STICH, Institute for Industrial Management at RWTH Aachen University, Germany
KLAUS WEHRLE, RWTH Aachen University, Germany

Supply chains form the backbone of modern economies and therefore require reliable information flows. In practice, however, supply chains face severe technical challenges, especially regarding security and privacy. In this work, we consolidate studies from supply chain management, information systems, and computer science from 2010–2021 in an interdisciplinary meta-survey to make this topic holistically accessible to interdisciplinary research. In particular, we identify a significant potential for computer scientists to remedy technical challenges and improve the robustness of information flows. We subsequently present a concise information flow-focused taxonomy for supply chains before discussing future research directions to provide possible entry points.

Funded by the Deutsche Forschungsgemeinschaft (DFG, German Research Foundation) under Germany's Excellence Strategy – EXC-2023 Internet of Production – 390621612 and the Alexander von Humboldt (AvH) Foundation.
Authors' addresses: J. Pennekamp, R. Matzutt, E. Vlad, S. J. J. Leemans, and K. Wehrle, Communication and Distributed Systems, RWTH Aachen University, Ahornstrasse 55, Aachen North Rhine-Westphalia, 52074, Germany; emails: {jan.pennekamp, roman.matzutt, vlad, klaus.wehrle}@comsys.rwth-aachen.de, s.leemans@bpm.rwth-aachen.de; C. Klinkmüller, BPMotion, 10 Gore Ave, Kirrawee, New South Wales, 2232, Australia; email: christopher.klinkmueller@bpmotion.com.au; L. Bader, M. Serror, and E. Wagner, Cyber Analysis & Defense, Fraunhofer FKIE, Fraunhoferstraße 20 Wachtberg, North Rhine-Westphalia, 53343, Germany; emails: {lennart.bader, martin.serror, eric.wagner}@fkie.fraunhofer.de; S. Malik, Software Systems Research Group, Data61 CSIRO, Level 5/13 Garden St Eveleigh New South Wales, 2015, Australia; email: sidra.malik@data61.csiro.au; M. Spiß, J. Rahn, and V. Stich, Production Management, Institute for Industrial Management at RWTH Aachen University, Campus-Boulevard 55, Aachen, North Rhine-Westphalia, 52074, Germany; emails: {maria.spiss, jessica.rahn, volker.stich}@fir.rwth-aachen.de; T. Gürpinar, Supply Chain Development and Strategy, Fraunhofer IML, Joseph-von-Fraunhofer-Straße 2-4, Dortmund, North Rhine-Westphalia, 44227, Germany; email: tan.guerpinar@iml.fraunhofer.de; S. S. Kanhere, School of Computer Science and Engineering, University of New South Wales, Engineering Rd Sydney, New South Wales, 2052, Australia; email: salil.kanhere@unsw.edu.au.

**32**

**ACM Reference format:**
Jan Pennekamp, Roman Matzutt, Christopher Klinkmüller, Lennart Bader, Martin Serror, Eric Wagner, Sidra Malik, Maria Spiß, Jessica Rahn, Tan Gürpinar, Eduard Vlad, Sander J. J. Leemans, Salil S. Kanhere, Volker Stich, and Klaus Wehrle. 2023. An Interdisciplinary Survey on Information Flows in Supply Chains. *ACM Comput. Surv.* 56, 2, Article 32 (September 2023), 38 pages.
https://doi.org/10.1145/3606693

## 1 INTRODUCTION

Supply chains connect organizations on a regional, national, and global level, enabling them to jointly manufacture products and offer services. The importance of well-functioning supply chains for both business and everyday life has been prominently demonstrated over the past years. Disruptions caused by climate change [94], COVID-19 [177], the Suez canal obstruction [93], or the Russian invasion of Ukraine [195] have a lasting impact on the production and distribution of goods and materials, resulting, e.g., in shortages of grain, fuel, pharmaceuticals, and semiconductors that affect manufacturers, consumers, and societies worldwide. As a result, organizations invest in strengthening the robustness and resilience of supply chains, i.e., the ability to maintain or quickly return to normal operation after disruptions [110]. Due to globally connected economies with complex supply chains, this endeavor requires holistic solutions, e.g., by establishing circular economies [13] or redesigning global production networks and manufactured products [110].

Data sharing and tighter collaborations between supply chain participants have been identified as prominent drivers for increasing the robustness and resilience of global supply chains [159, 202]. In fact, the establishment of reliable communication channels has long been recognized as a critical success factor for managing supply chains [1] as well as the business processes the supply chains run on top of [143]; in the face of today's challenges, this requirement only gains further importance.

For example, production planning and control requires organizations to exchange accurate delivery estimates and demand forecasts, e.g., to avoid the bullwhip effect [38]. Similarly, to identify and react to unforeseen circumstances quickly, real-time monitoring systems must be established, and they must be able to collect information from the entire supply chain [92, 100, 154, 214]. Lastly, provenance information is required to recall defunct products and to detect their origins by tracing them upstream in supply chains [56, 129].

As a consequence, ***digital supply chains*** (***DSCs***), which center around the automated and digitized information exchange between organizations, have emerged and gained further traction in recent years [27, 151, 190]. This notable shift increasingly connects traditional supply chain management to IT-based solutions. However, related work so far has primarily been driven by supply chain experts, who may not be aware of state-of-the-art solutions in potentially beneficial IT-related research. Missing this opportunity could lead to unused potentials in both supply chain research and real-world deployments of DSCs.

While organizational barriers, such as a missing sense of urgency or costs [4], are present, organizations also face severe technological challenges, e.g., related to IT security and privacy when setting up the required communication infrastructure [103, 181], for which they lack the corresponding competence and skills [66].

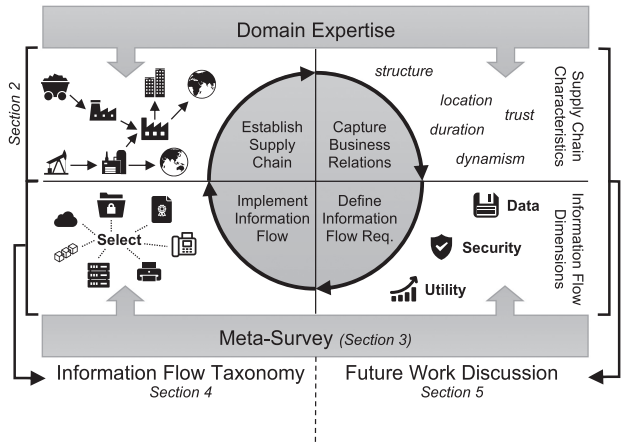

Fig. 1. The foundation and methodology of our survey article.

In light of the transition from traditional to digitized supply chains, in this article, *we investigate the extent and building blocks of technical contributions that promise reliable information flows.* Considering that the academic literature on supply chain management is vast and diverse, we first have to gain an overview of which computer science-related ideas are already well-accepted in the supply-chain community. Accordingly, we center our efforts on analyzing to which extent these ideas are captured in survey articles and recent research articles from the supply-chain domain. Overall, we find that survey articles in this domain focus on quantitative aspects (articles per year, articles by region or venue, used methods, and others) for specific factors only. Despite the acclaim of their importance, more emphasis should be put on the essential, crucial information flows, their underlying data communication, and the means and concepts to implement them. Furthermore, we observe that corresponding (academic) solutions are either (1) not on-point (i.e., too general), (2) overly use case-specific, (3) with a strictly economic focus only, or (4) not properly evaluated. Consequently, the latest advances in IT-based solutions have little influence on real-world deployments and the evolution of supply chains. In conclusion, we find that computer science research has widely neglected to establish reliable information flows in supply chains, thus impeding sustainable technical solutions from a computer science perspective.

With our work, we systematically survey the current state of reaching reliability for information (flows) as a critical building block for DSCs to incentivize further collaboration between computer scientists and supply-chain experts in the future. To this end, we qualitatively evaluate to which extent technological solutions and relevant security needs for establishing (reliable) information flows are already prevalent in the context of supply chains.

Furthermore, we follow a holistic approach as proposed by Peng et al. [127]; that is, we comprehensively analyze the impact of the respective technology on supply chains. Generally, we adopt the approach in Figure 1: Based on domain knowledge in foundational supply chain literature, we summarize key supply chain characteristics as a first step. These characteristics enable us to systematically discuss the nature of a specific supply chain and, in particular, the business relations within it. We then focus on information flows that need to be established to support use cases such as production planning or tracking and tracing. Here, we conduct a qualitative meta-survey, i.e., we specifically target existing survey articles to analyze the state of the art. Our approach has two distinct advantages. First, focusing on survey articles allows us to efficiently cover a vast research corpus of thousands of research articles. Second, this approach enables us to identify which general ideas are well-known and accepted in the respective research community. We argue that covered articles and ideas are likely to have a certain level of visibility and potentially even impact on future developments.

Our corresponding meta-survey covers the period 2010–2021, i.e., we capture developments of more than one decade. Based on our meta-survey, we derive an information flow taxonomy. The objective of this step is to systematically describe requirements on the information flows along three dimensions: (1) *data* (the nature of the data that needs to be exchanged), (2) *security* (aspects focusing on the protection of exchanged information), and (3) *utility* (aspects related to the quality of the information flow). In a subsequent step, we then utilize the supply chain characteristics and our novel taxonomy to discuss the current state of technical solutions for implementing (new) information flows within specific supply chains and eventually discuss future research directions.

**Contributions.** Our contributions are threefold. First, we conduct a systematic literature review to provide a comprehensive overview of information flows in supply chains, revealing the neglected view of the full information lifecycle. Second, we consolidate the discussions of different strings of research by (a) formalizing the characteristics of supply chains and (b) deriving a taxonomy on information flow-focused supply chain research to facilitate interdisciplinary exchange. Third, based on our findings, we derive the need for in-depth interdisciplinary research, with an emphasis on the reliability of information flows in supply chains. This work thus lays the foundation to make this interdisciplinary topic holistically accessible to computer scientists and supply chain experts.

**Organization.** The remainder of this article is structured as follows. Section 2 introduces fundamental background on supply chains (including characteristics to describe business relations within). Section 3 presents our meta-survey, including our research questions, the review approach, an overview of the identified publications, and a brief discussion of the respective publications. Based on our survey, Section 4 introduces our supply chain information flow taxonomy, reviews existing terms that characterize information flows against our taxonomy, and exemplifies our taxonomy using common supply chain use cases. Section 5 discusses currently (mostly) untapped research directions and how they are required for a successful future supply chain management on a more general level, covering the intersection of supply chain and computer science research. Finally, Section 6 concludes this survey article.

## 2 MOTIVATION AND BACKGROUND

As a foundation for our analysis of information flows, we first establish a common basis concerning supply chains from a business perspective that is also comprehensible for other domains. To this end, in Section 2.1, we give a brief overview of **supply chain management (SCM)**. Subsequently, in Section 2.2, we introduce the dimensions of digital SCM and we describe its relation with business process management in Section 2.3. Moving toward information flows along supply chains, we highlight well-established use cases in Section 2.4. Based on this overview, Section 2.5 then discusses SCM and its information flows from a computer science perspective, i.e., we present analogies to well-known computer science concepts to ease the comprehension of this business-focused view.

### 2.1 Supply Chain Management (SCM)

Traditionally, supply chains are primarily concerned with the flow of (physical) products. To improve cross-company business processes, supply chains also increasingly implement information flows between their participants. In contrast to mostly unidirectional product flows, information flows are implemented both upstream and downstream (cf. Figure 2), and even detached from the original supply chain structure. More importantly though, information flows are not necessarily limited to a single hop, i.e., they might cover multiple hops (vertical collaboration), e.g., between

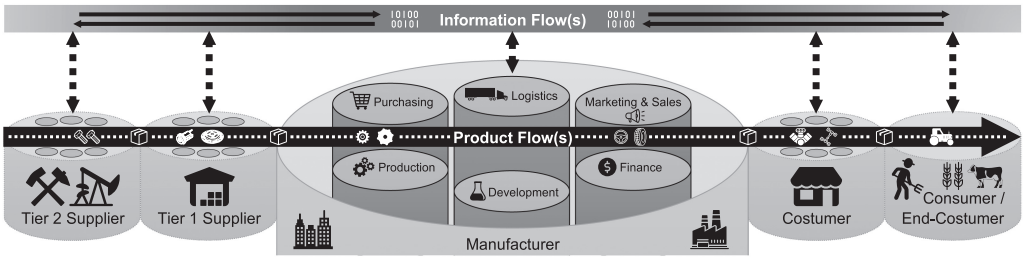

Fig. 2. Product and information flows concern a number of stakeholders along the supply chain (based on [89]). While product flows are unidirectional, information flows are established in both directions and potentially cover multiple hops. As such, data communication is the backbone of modern supply chains.

tier-1 suppliers and consumers [35, 84], different branches, e.g., multiple customers of a single supplier, or even structures that exceed the scope of a supply chain (horizontal collaboration) [16]. In fact, numerous actors are involved in supply chains, and the complexity increases constantly [55, 111, 123, 145]. Extending on Figure 2, distributors, retailers, commodity corporations, and many other actors can be involved in complex supply chain networks [111].

SCM coordinates and optimizes the material, information, and financial flows throughout the supply chain [108, 160] while aiming at the maximization of value creation [196]. The affected supply chain processes range from raw material extraction to delivering the manufactured product while considering consumer relationships, customer service and demands, order fulfillment, manufacturing flow, procurement, and product development, as well as commercialization and respective return processes [33]. The fundamental idea of SCM is based on the belief that efficiency can be improved through information sharing, joint planning across suppliers and customers, and other collaborations [21].

The effectiveness of SCM also depends on real-time processing of information, process alignment (supplier integration), (aligned) decision-making [197], and synchronized financial flows [108]. In particular, order management, production planning, data management, and tracking can counteract the efficiency decline of the supply chain [73]. As such, SCM enables the systematic planning, management, and coordination of supply chains. However, only when implemented effectively, SCM helps its diverse actors to monitor the supply chain in real-time to promote its responsiveness [164] as well as resilience [98]. With the increasing awareness of SCM success factors, stakeholders are gradually beginning to provide and source more information within supply chains [208].

Overall, SCM concerns both short-term and long-term decisions and strategies of a multitude of stakeholders in supply chains to handle product, financial, and information flows.

### 2.2 Dimensions of Digital SCM

With the past and still ongoing developments of supply chains toward increasingly integrated suppliers, and more dynamic business relationships, the importance of digitized supply chains, their management, and the integration of novel technologies to support (data) communication is significantly increasing [61, 111, 209]. Thus, overall, supply chains transitioned to complex and global networks, encompassing a large number of stakeholders. Moreover, the reduction of trade barriers and greater interconnection of additional supply chain partners are driving the scope and growth of a global economy.

*2.2.1 Supply Chain Characteristics.* Conceptually, various features that determine the business relationship influence the characteristics of supply chains. The *structural complexity* of supply chains is influenced by the system size, the degree of order (linkage), and the categories of elements [31]. Here, the elements include the different members of the supply chains (e.g., suppliers, manufacturers, distributors) as well as the information, product, and financial flows [31, 163]. The supply chain, modeled as a graph or network, is further characterized by the path length (average number of actors or tiers that must be traversed between any two actors), the connectivity distribution (average number of connections possessed by each node in the network) and the clustering coefficient (expresses network transitivity, i.e., the average probability of two neighboring nodes that are connected to a given local node are also connected to each other) [65]. Moreover, different network structures (small-world, scale-free, community, and hierarchical) exist in practice [201].

Depending on the *geographic location* of supply chain partners, we also gradually distinguish local and global supply chains [101]. Additionally, the *duration of collaboration* can range from short-term to long-term, and the relationship can be coordinated or a pure exchange relationship [165]. When looking at the *dynamism* in the choice of business partners, supply chains can be static (partners are relatively stable) or dynamic (partners vary depending on the market opportunity) [87]. Thus, based on the dynamism and the background of relationships, *trust* is a crucial factor as well [60].

Fawcett et al. [46] distinguish four levels of trust (limited trust, transactional trust, relational trust, and collaborative trust). With limited trust, the focus on the relationship lies in obtaining the lowest short-term cost at a fixed quality level. Transactional trust forms arms-length relationships, while in relationships based on relational trust, collaborative behavior increases. Collaborative trust is a close relationship focused on mutual success, joint planning and problem-solving, and increased competitiveness of the whole supply chain.

Especially in large supply chain networks, companies are concerned with safeguarding sensitive information and trade secrets [28, 96, 118]. Moreover, reservations regarding new technologies (e.g., blockchain technology) can impair the establishment of information flows [61, 127, 198], and, thus, negatively affect SCM. Therefore, approaches to securely exchange information along supply chains, irrespective of the supply chain's individual characteristics, are critical.

*2.2.2 Types of Information Flows.* In light of developments toward global communication, interconnectivity, and integration, the potential for new business models arises: Data has become a crucial asset for the creation of value in companies' operations [52]. The large amount of data leads to disruptions of established value creation structures as well as traditional business models and offers opportunities for innovative products and services. However, these new data-driven innovations cannot be advanced by a single stakeholder [145]. Instead, increasingly interconnected supply chains lead to the combination, enrichment, and sharing of various data sources from different actors in cross-industry data ecosystems [52].

For traditional SCM (cf. Section 2.1), we can differentiate between repeated and one-time information flows. While the former is linked to subscribed events, the latter is usually present for specifically exchanging product information.

**Repeated Flows.** Such flows increase the transparency of the current status or other relevant information, for example, with the goal to reduce uncertainties or improve reliability during the planning process by extending the amount of available information (cf. bullwhip effect [91]). Corresponding information flows entail a quicker availability of information. Thus, they might allow companies to improve their supply chain resilience, deal with disruptions, and improve their flexibility in general. They further contribute to reducing the latency until a decision has been made, as the reduced data latency allows companies to analyze reported events more quickly [216]. Consequently, after a taken action, the overall time until its measures show effect is reduced.

**One-Time Sharing.** These flows primarily improve the transparency regarding a specific product and its quality. They also address issues with documentation (e.g., fair-trade products, sustainability, or authenticity) and traceability along the supply chain in general. However, corresponding information flows can also trigger actions that cause repeated flows. For example, minor production deviations might be acceptable for one customer but not for the originally intended recipient [189].

### 2.3 Business Process Management and Supply Chain Management

Digitized SCM and the interaction between different supply chain actors, with respective information and financial flows, are also closely related to business processes. **Business process management (BPM)** can support the management of such processes. In particular, BPM oversees how work is performed in an organization to achieve consistent outcomes [43]. As such, BPM considers the flow of work (control flow), the flow of information and physical artifacts (data perspective), and who performs particular tasks (resource perspective). Success in SCM and BPM goes hand in hand: internal processes that are developed with supply chain members in mind have been shown to have lower costs and satisfy service requirements better [143]. Moreover, both BPM and SCM are vital for performance improvements and competitiveness [143]. Focusing on BPM practices also helps to support collaborative activities with supply chain partners of an organization [169].

Traditionally, BPM has focused on knowledge-intensive work processes by coordinating work and information flows *within* an organization [18]. In Figure 2, BPM would consider all processes within the manufacturer, spanning purchasing, logistics, marketing & sales, finance, development, and production. For instance, a business process may start as an engagement with marketing, then lead to a sale for which parts are purchased, and finally, the ordered products are manufactured and delivered by logistics. Business processes are well supported by information systems: A **BPM system (BPMS)** can be given a process model and will execute it for each incoming case by distributing work items amongst human and robotic workers, thereby interacting with auxiliary digital systems [2].

While BPM mostly focuses on intra-organizational aspects, recently, inter-organizational BPM emerged to support collaboration between organizations [143]. For instance, choreography diagrams of the BPMN standard [125] provide a means to describe the communication between organizations that collaborate to achieve positive outcomes in an orchestrator-less setting. Choreography diagrams form a bridge between inter-organizational BPM and research on protocols, such as validation of properties using model checking. However, the use of such diagrams for supply chain analysis is challenged by the coarseness of the modeled communication. Moreover, BPMSs do not adequately support choreographies or processes that span multiple organizations [42], with only a few exceptions [3]. On account of describing the flow of sensitive information, BPM also entails essential security needs without focusing on them.

In BPM, analysis and improvement of existing business processes play a significant role [43]. To support these steps, process mining offers automated tools to gain insights into running processes from recorded event data [187]. The need for automated analysis techniques for inter-organizational process mining has been identified [187] and initial work has been performed [183, 188]. While corresponding concepts that deal with the flow of information have been applied to supply chains (e.g., process mining techniques) [25], they do not consider technical aspects of (reliable) information flows, which we focus on in this article. Thus, despite the outlined interplay between BPM and SCM for supply chains in real-world deployments, in the following, we place our emphasis on the impact of digital SCM on information flows. Accordingly, we continue with an introduction of the most common inter-organizational SCM use cases in modern supply chain networks.

### 2.4 Common Use Cases in Supply Chains

We now introduce the motivation as well as the workflows of typical use cases along with their requirements regarding information flows in the context of supply chains and SCM.

**UC1: Collaborative Planning.** A well-known problem of supply chains with multiple hops is the bullwhip effect [91]. To allow for improved production planning and enhanced demand forecasts, transparency along the entire supply chain is needed, i.e., companies should share their and their suppliers' changes in demand with their customers [92]. This workflow would enable a simple adaption of the capacity planning based on capacity utilization within the supply chain [100]. However, sharing these business insights might also provide competitors with valuable insights [184]. Regardless, several approaches and their corresponding information flows are increasingly prevalent in today's supply chains. Exemplary collaboration concepts include **vendor-managed inventory** (**VMI**), **collaborative planning, forecasting, and replenishment** (**CPFR**), and **just-in-sequence** (**JiS**) inventory management to realize a **just-in-time** (**JiT**) production [68, 92, 174].

**UC2: Supply Chain Design.** Maintaining a global supply chain network and especially bootstrapping new business relationships is a significant challenge for companies [62, 141]. For example, manufacturing a new product might require a completely new set of suppliers. In light of custom production, reacting to customer change requests is a crucial aspect of manufacturing. In such a setting, a manufacturer is interested in suitable business partners, while potential suppliers do not necessarily want to globally announce their production and delivery capabilities [130, 131]. Regardless, reliable on-demand information sharing on available capabilities could establish a fairer and more competitive market with direct implications on the design and maintenance of supply chains [133].

**UC3: Tracking – Real-Time Monitoring.** To primarily anticipate problems as early as possible, manufacturers are interested in full visibility of the upstream and downstream activities [56, 141]. When transported, goods are often handled by many different parties, including lead logistics, carriers, shipping lines, ports, airports, and customs. These parties are independent, and their cooperation is often limited to a single hop which significantly challenges the uninterrupted monitoring [92]. Overall, this use case is not limited to location information only. It can also cover condition monitoring, e.g., whether a cold chain was intact during transit [17, 155]. Besides, tracking data helps in improving delivery date predictions (ETAs) and time slot management [15, 142].

**UC4: Tracing – Handling Faults.** Likewise, when identifying issues with products after the fact, manufacturers are interested in tracing the product along the supply chain to the current user (i.e., customer) while identifying used components, tools, and products as well as the respective suppliers [24, 149]. Such an approach allows them to be informed, for example, about reduced product lifetimes or improper operational reliability [185]. Moreover, they can inquire about their products' usage data to obtain a better understanding of the extent and severity of the fault at hand. If needed, they can also instruct a product recall, a practice that is common for food supply chains or in the event of safety-critical failures (cf. automotive or aviation industry). Established data sharing reduces follow-up costs and application latency until the measures show their effect.

**UC5: Tracing – Sourcing Faults.** Conversely, a customer can also identify an issue with a product, potentially in a specific subcomponent only, i.e., identifying the root cause is of interest. To this end, tracing the product and its components backward through the supply chain is a suitable approach [147]. Successfully pinpointing the failure's origin also benefits other companies with similar or identical products in use, effectively triggering a tracing process [60].

**UC6: Tracing – Validation.** As specified by customers or by law, supply chains need to abide by regulations and contracts (e.g., the supply chain act [26]). For example, auditors might want to verify that all regulations are followed [147]. Thus, complete, unmodified, and accurate historical information about the activities and production processes in the supply chain is needed. Such verifications can also occur through third-party regulators or certifiers, e.g., to look at sustainably sourced products, organic food items, or authentic origins of diamonds. Similarly, digital certificates regarding the product's quality can be of value. Especially with pharmaceutical products [40, 59] or art [148], staying clear of counterfeit products is a crucial requirement. Even more, the benefits of tracing also translate to end-customers who increasingly care whether their products are ethically and sustainably sourced [158].

**UC7: Critical Infrastructures.** Especially for international supply chains, goods flow through critical infrastructures, such as ports or airports. There, operators interact with different logistics companies. Often, information about the freight that results in actions by the operator, e.g., storage, pickup, and safety and custom inspections of freight, is provided by the lead logistics but stems from other, original shipping companies. Most prominently, the effect of unreliable information in this context has been demonstrated in the Antwerp container port hack, where information about containers was altered to bypass inspections and to smuggle drugs [162]. Consequently, the impact of unreliable information in these infrastructures can be enormous on both society and economy.

**UC8: Product Information.** Information on products is not only relevant for production planning (e.g., for logistics and inventory), but it also allows companies to adjust their manufacturing accordingly [120, 172]. Thereby, manufacturers can, for example, react to slight deviations or identify critical issues early on. Thus, sharing detailed information about an individual part from the supplier to subsequent customers, even over multiple hops, can entail significant process improvements and benefits. However, given its sensitivity and the potentially adverse effects for suppliers (e.g., by willingly reporting on deviations), this practice is not yet widely established [203].

Each use case expresses specific needs for the exchange of information, i.e., the respective information flows vary even across different implementations of the same use case due to the varying supply chain characteristics (cf. Section 2.2.1). However, in the end, they all build upon access to reliable information [37]. Thus, in the next sections, we specifically look into challenges regarding the design and development of reliable digital information sharing in supply chains.

### 2.5 A Technical Perspective on Information Flows

Before looking at these information flows in more detail, we first want to improve the comprehension of SCM and its information flows for computer scientists by introducing two analogies: First, we compare them to the network stack, and, second, we match decision processes within (global) supply chains to control loops in **cyber-physical systems** (**CPSs**).

**Different Layers Interacting.** As for the network stack in communication systems, data sensed in supply chains is also passed through different logical layers until decisions are made, as we illustrate in Figure 3(a). Here, the upper two layers depend on (reliable) data for their decision-making. We consider the exact physical handling as well as the specific decision-making, i.e., SCM planning algorithms, as out of scope for this article. Instead, we focus on the flow of information and the associated data processing.

**Supply Chain Decisions.** As we model in Figure 3(c), we can identify a decision loop within supply chains, which sources information during the decision-making that is partly based on suppliers forwarding information (e.g., on products and shipments) to allow them to adjust their (local) processes accordingly. Here, most of the decision loop steps are significantly influenced by external (global) actions and (sensed) information. Thus, it requires reliable information.

*Physical Process.* This step cannot be digitized as it concerns real-world shipments. Its events are the main source of information for decision-making in the context of SCM.

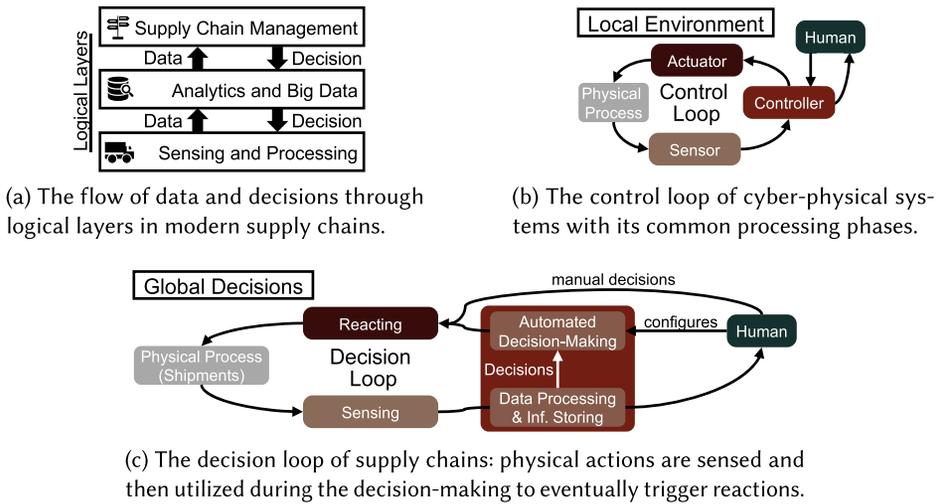

(a) The flow of data and decisions through logical layers in modern supply chains.

(b) The control loop of cyber-physical systems with its common processing phases.

(c) The decision loop of supply chains: physical actions are sensed and then utilized during the decision-making to eventually trigger reactions.

Fig. 3. Analogies from the domain of computer science to illustrate the decision-making in supply chains.

*Sensing.* The sensing step is needed to capture and digitize the information of physical processes and concerns all kinds of information, e.g., shipment status, environmental conditions, or processing of goods along the supply chain.

*Decision-Making.* Based on the available information, companies can make decisions to adjust shipments or production planning (e.g., sourcing components from another supplier). Depending on the information quality and magnitude of the decision, this step is either human-managed or automated. In the future, we expect increased automation following AI-driven advances, potentially with the need for a human operator to simply confirm suggested changes.

*Reacting.* This step basically orchestrates any required changes that follow from the decision-making to the physical processes. Consequently, this step manipulates real-world shipments and/or production processes accordingly.

**CPS Control Loop.** Our suggested supply chain model is an analogy to the CPS model on control loops [179], which features similar phases and challenges (cf. Figure 3(b)), i.e., sensing, controlling, and acting. However, in contrast to digitized supply chains, CPS control loops usually source their information locally. Thus, a single control cycle is much quicker than in supply chains, where we consider a global view with potentially multi-hop information flows from different stakeholders. Thus, our supply chain model is much broader than the traditional CPS model.

Concerning the reliability challenges of information flows, transferring security and privacy analyses from the CPS model to supply chains could be a wise idea to capture all relevant aspects within global supply chain environments.

## 3 A META-SURVEY ON INFORMATION FLOWS IN SUPPLY CHAINS

This article aims at making the topic of information flows in supply chains readily accessible to interdisciplinary research and experts from different domains. Therefore, we conduct a **systematic literature review (SLR)** targeting the current state of (technical) research on information flows in supply chains. For this work, we focus on a qualitative survey and refrain from presenting a quantitative analysis as it would (1) introduce an inherent bias based on the underlying surveys' inclusion criteria, (2) add little value w.r.t. the goals of our article, and (3) lack depth and insights due to the expressed oversight of compelling and profound research to date (cf. Section 4; we refer to relevant articles when presenting our taxonomy). Our corresponding presentation is as follows.

First, in Section 3.1, we outline the primary research questions for our SLR. Subsequently, in Section 3.2, we detail our SLR methodology, including potential limitations. Then, we present statistics on the SLR process (Section 3.3) and general characteristics to establish a high-level overview of relevant work in the area (Section 3.4). Finally, in Section 3.5, we discuss the main findings of our SLR before concluding this section with takeaways on the evolution in Section 3.6.

### 3.1 Research Questions

With our survey, we assess the current state of information flows in the context of supply chains while also providing a detailed, technical view of the underlying concepts. This approach thus provides a viewpoint that does not primarily focus on the business perspective and primarily focuses on two aspects:

(1) What are the limitations of existing approaches implementing reliable information flows in supply chains?
(2) How can general requirements for information flows in supply chains be systematically described and grouped?

Led by these research questions, we thoroughly conducted an SLR. Before presenting and discussing our findings, we first detail our methodology in the following.

### 3.2 Survey Methodology

Research in the area of supply chains is characterized by an exceptional amount and wide range of work. To thoroughly capture this domain, we focus our SLR on existing *literature surveys* (no questionnaire-based surveys), which cover more than one decade of research and developments in the area. Furthermore, we resort to relatively general keywords to identify relevant articles. The benefits of this approach are twofold. First, we get a curated overview of the most important research directions in the field without excluding potentially relevant articles too early. Second, individual proposals of single articles do not bias our overview and understanding since the considered survey articles are bound to contextualize the articles' impact within the covered research area. Not considering individual proposals in our methodology could also be seen as a limitation hindering the immediate transfer of existing solutions. However, we argue that this work should lay the foundation for a sustainable interdisciplinary approach by identifying reliable and commonly known technological building blocks instead of individually recommending specific and possibly outdated or obsolete solutions.

Concerning our methodology, we adapt distinct best practices [23, 86, 135, 136]: We use *Parsifal* [48] to broadly identify potentially relevant literature from Scopus and **Web of Science (WoS)**, published in 2010+. Namely, we started with all articles matching the following query either in title, abstract, or keywords: *supply chain* ∧ (*information* ∨ *data*) ∧ (*literature* ∧ (*survey* ∨ *review*)). After a broad initial filtering, e.g., to remove duplicates, we conducted two increasingly refined rounds of content-based filtering, i.e., we first screened the titles and abstracts and then assessed the introductions and conclusions of the respectively remaining articles. Finally, we systematically analyzed all remaining articles in preparation for this survey article.

### 3.3 Statistics of Our Conducted Meta-Survey

On October 4 2021, we extracted 2,708 articles from Scopus and 2,769 articles from WoS matching our search query. During the initial filtering, we removed duplicates (1,306), full proceedings (25), and non-English articles (5). After this step, a total of 4,141 articles remained for further consideration.

Table 1. For our interdisciplinary meta-survey, we eventually considered 70 survey papers

| Year | # | Papers | Year | # | Papers |
|---|---|---|---|---|---|
| 2010 | 4 | [10, 57, 102, 156] | 2017 | 2 | [157, 213] |
| 2011 | 5 | [8, 134, 161, 192, 212] | 2018 | 4 | [9, 30, 175, 193] |
| 2012 | 2 | [117, 210] | 2019 | 17 | [20, 32, 36, 44, 51, 58, 70, 76, 95, 107, 121, 122, 144, 166, 167, 199, 206] |
| 2013 | 6 | [19, 53, 78, 82, 178, 207] | | | |
| 2014 | 9 | [7, 69, 72, 79, 81, 85, 113, 137, 152] | 2020 | 6 | [29, 34, 67, 116, 119, 215] |
| 2015 | 1 | [170] | 2021 | 11 | [5, 12, 45, 97, 99, 109, 112, 115, 171, 182, 186] |
| 2016 | 3 | [77, 150, 204] | | | |

We then proceeded to screen titles and abstracts of the remaining articles. Here, we excluded 3,786 articles overall: At least 1,268 articles were deemed off-topic (i.e., research not primarily concerned with supply chains), at least 1,121 articles had orthogonal research questions (they covered supply chains or logistics more remotely), and at least 217 articles reported on survey types unsuitable for a meta-survey (e.g., they were based on expert interviews or questionnaires). After this screening, a total of 355 articles remained eligible.

Subsequently, we obtained the available full texts of the remaining articles, read their introductions and conclusions, and assessed the quality of the articles' publishing outlets. Here, we identified 175 articles to be irrelevant to our meta-survey (we were unable to access the full text of 17 articles), and we excluded 101 articles deemed only "partially relevant," i.e., they fell below our scoring threshold in Parsifal. Furthermore, we identified one article that had been retracted previously; hence, we excluded it as well. Thus, as a basis for our meta-survey, we obtained 79 surveys that were either highly related to research on (reliable) information flows in supply chains or partly related but published at high-impact venues.

During the final reading of these 79 articles, we further excluded 9 ineligible articles that could only be excluded based on the full text, leaving us with 70 articles in total (Table 1), whereby most of them have a primary background in the engineering, computer science, or business domains.

### 3.4 Content Systemization

Now, we give a general overview of the fully read articles of our SLR. Most notably, our meta-survey revealed a large body of especially extensive surveys (including SLRs). The extensiveness of these surveys (hundreds [9] to thousands [175] of considered articles) underpins the availability of a large but insufficiently structured body of knowledge in this area. We visualize the time-wise distribution of the surveyed articles per domain along with the publication years of articles cited by these surveys as an indicator for research interest in Figure 4.

Content-wise, the surveys usually emphasize one specific topic for a detailed analysis (e.g., value of information [193] or supply chain resilience [109]). Overall, these surveys mainly approach and evaluate information flows in supply chains from a business perspective, e.g., with a focus on the implications on BPM [10, 29]. The most prominent research area that utilizes shared information (cf. Figure 3) is "smart" decision-making (e.g., [30, 137, 171, 213]). However, in light of our focus on information flows (cf. Section 3.1), we consider this area mostly out of scope. Other particularly business-driven topics are the bullwhip effect (e.g., [32, 53, 67, 122]), logistics management (e.g., [19, 30, 115, 119, 134, 167, 178]), and the quality of business data (e.g., [9, 77, 157, 175, 193, 212]).

Still, information sharing seemingly has become more and more relevant to supply chain experts in recent years, as indicated by the increasing numbers of relevant articles (cf. Table 1). The studied surveys cover a wide range of years that even date back to 1993 [213]. Use cases (cf. Section 2.4) related to tracking and tracing (e.g., [12, 19, 51, 70, 156, 215]) are studied especially frequently. Contrarily, we noticed that technical discussions were either lacking or underrepresented in the studied literature. While terms such as information technology, security, or interoperability are mentioned, they are not elaborated on, i.e., *technical aspects are considered crucial, but they mostly remain unexplored*.

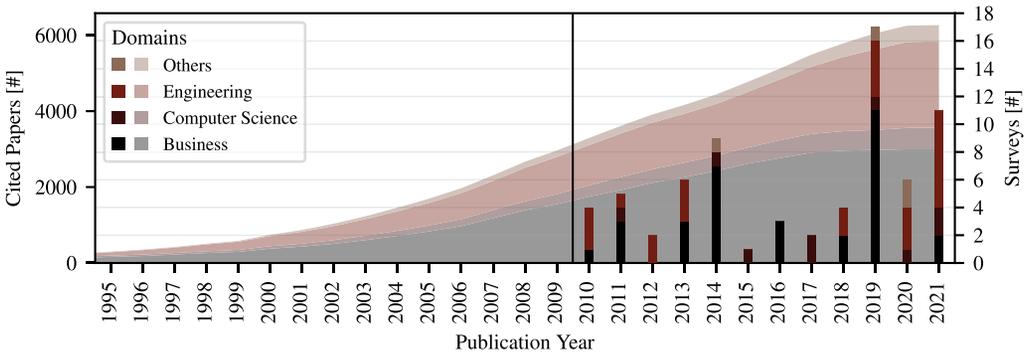

Fig. 4. For our interdisciplinary meta-survey, we considered articles from three primary domains (as derived from their publication venues), namely, engineering, computer science, and business. We visualize their distribution across the surveyed years. Furthermore, we indicate the cumulative number of articles cited by these surveys to provide an indicator for research interest across the respective domains and years.

As supply chains have gained universal relevance, surveys tend to focus on specific sectors. A larger number of articles covers food supply chains (e.g., [19, 30, 34, 82, 99, 116, 152, 213]), including the branches on cold chains and agricultural supply chains. In addition, we also discover targeted areas, such as pharmaceutical supply chains [36] (a focus on certification and origin tracing), steel processing [215] (product identification and traceability), or agile manufacturing [58] (complex supply chain networks). Overall, past literature covers applications and use cases in various domains.

Moving toward more technical aspects, we want to highlight that nearly all surveys touch upon a number of technology-related concepts and paradigms. Well-known concepts, such as the Internet of Things, cloud computing, Big Data, machine learning, or artificial intelligence, are frequently mentioned in these works. In particular, a specific emphasis is put on RFID, constituting the link between physical flows and digital supply chains. Likewise, technologies such as electronic product codes, bar codes, and QR codes are also present. When looking at the communication, we observe that (1) either no details on how the information is exchanged are given at all, (2) old and inflexible technologies are suggested or in use (e.g., fax [30, 72, 85, 163, 212], SMS [30], or email [8, 72, 85, 212]), or (3) a technology is named without giving details. Despite the different research domains of the surveyed articles, we observe a significant cross-domain coverage, i.e., although not always discussed in great detail, many surveys recognize the importance of other domains' aspects.

Overall, the technical perspective, which would assume the task of augmenting these crucial discussions and developments with the required background, is severely underrepresented. Next, we discuss respectively arising challenges in more detail.

### 3.5 Discussion on Data Sharing and Information Flows

Now, we discuss the most prominent aspects of information flows and collaboration raised by our meta-survey. This section provides a solid overview of the current state of the art in research.

**Motivation.** Collaboration and data sharing can greatly improve the supply chain performance as well as resilience [32, 45, 166, 204]. Besides, the dynamism in business relationships can be improved [12] as needed to account for customer requests and design specifications [58]. So far, only restricted data sharing is already implemented. Consequently, several challenges remain on the road toward globally collaborating supply chain networks (e.g., [112]).

**Enabler.** In this regard, the quality or value of information is a crucial aspect [77, 193]. Only when effectively integrating the information sharing in SCM, the decision-making, and in turn, performance and resilience can be improved. In this context, standardization and governance are further needed to reliably enable information flows [30, 70, 76, 152].

**Information Flows.** As the establishment of information flows becomes a necessity [32, 122], further research is required to improve the coordination [171], multi-hop collaboration [107], and dynamism of supply chains [9]. Especially, recent initiatives (e.g., Internet of Production [22, 132] or Physical Internet [14, 114]) with their wide range of benefits mandate the large-scale sharing of information, and, as such, also the implementation of information flows [12, 115].

**Concerns.** Unfortunately, the evolution toward extensive data sharing with many information flows also spawns diverse concerns. Apart from the risks of (un-)intentionally leaking sensitive information [9, 32] or drowning in information through oversharing [204], trust and its establishment constitute another major challenge [32, 116, 199]. Hence, future work should take them seriously and address these concerns by coming up with appropriate and scalable technical solutions.

**Threats.** Apart from these concerns by stakeholders, the literature also discusses a number of other threats, with (cyber) attacks being mentioned most frequently (e.g., [19, 44, 182]). Additionally, scalability and regulatory questions are notable challenges [99]. However, the primary focus is on internal threats [32], i.e., information security [170], including access control [9], is of significant relevance. To our understanding, corresponding countermeasures and concepts are rarely analyzed in light of their universal applicability.

**Building Blocks.** Finally, to address these concerns and threats and to allow for a successful implementation of information sharing, secure building blocks are needed. Especially, their international application [107], general interoperability [20], and cross-domain applicability (including different industry sectors) [215] remain challenging to date. Many recent surveys identify blockchain technology as a promising solution [74, 97, 166, 199]. However, so far, many solutions remain at a prototypical level [44], and concerns about their maturity exist [116]. These challenges are understandable as an application of blockchain technology in the context of supply chains is still considered to be in its infancy [119]. As such, privacy, throughput, and scalability issues should be resolved in the future [12, 71]. In Section 4.4, we will provide a more elaborate overview of currently proposed solutions.

Moving on, we will conclude our meta-survey before presenting our derived information flow taxonomy in Section 4.

### 3.6 Main Implications for the Evolution of Supply Chains

The future evolution of supply chains will be challenging and very interesting to observe, for example, the implications and developments following increasingly implemented coopetition [107]. Along with the increasing number of articles on the Industrial IoT, the number of published surveys on supply chains further seems to indicate an increasing activity in the field (cf. Table 1).

However, evaluating the impact of approaches is still difficult as a bridge to real-world use and deployments is missing so far. A re-iteration of research questions over time is mostly missing. According to our meta-survey, real-world evolution is rarely studied (only [32]). Besides, due to the focus on specific use cases, we should question their universality, not only in the context of information flows but also more generally, as it is difficult for computer scientists to draw meaningful conclusions without feedback from supply chain experts. In our view, any technically driven research effort is severely hindered by the insufficient exchange of open challenges, needs, and goals. On a more general note, we attribute the observed lack of technical details to the authors' backgrounds and their primary focus on business implications, i.e., a computer science (or at least interdisciplinary) perspective is missing for most work.

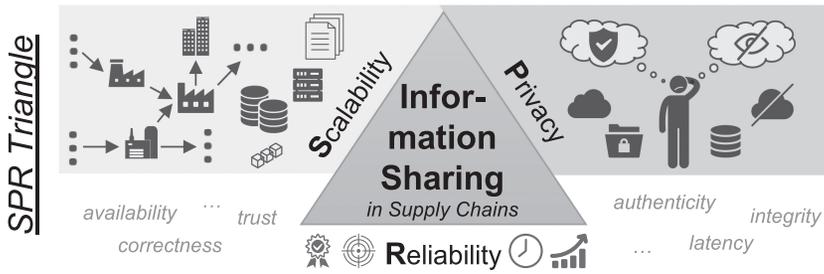

Fig. 5. Simultaneously fulfilling the three presented aspects for information sharing in practice is non-trivial.

Yet, we observe an inherent tension between the often-requested general properties for information sharing in supply chains on a technical level, which is reminiscent of another well-known result from computer science. Namely, the CAP theorem [47] captures the inherent tradeoffs between consistency, availability, and partition tolerance in any distributed storage system. Likewise, we argue that **S**calability, **P**rivacy, and **R**eliability are unlikely to be satisfiable simultaneously with reasonable effort when sharing information along supply chains. These tensions are further expressed in our corresponding *SPR triangle* in Figure 5.

Intuitively, focusing on any particular property can imply limitations on the other two. Measures to ensure reliability and privacy both add processing costs, which negatively impact the performance and ultimately the scalability when sharing information. Conversely, maximizing the scalability in the absence of major technical improvements implies compromises on either reliability, privacy, or both. Finally, achieving full data privacy and full reliability is mutually exclusive as privacy implies confidential data exchanges and full reliability involves ensuring that information flows are transparent, e.g., to facilitate the higher-level decision-making processes (cf. Section 2.5).

Building upon this intermediate takeaway, we next set out to formalize information flows and their technical properties to allow for a better understanding of these three aspects in practice.

## 4 A TAXONOMY FOR SUPPLY CHAINS

We now present our information flow-centric taxonomy for supply chains from an interdisciplinary perspective based on our SLR. We formalize our taxonomy in Section 4.1. Specifically, we group ten properties into three dimensions to structure and simplify its application. In Section 4.2, we present our taxonomy and its properties in light of the common supply chain use cases we detailed in Section 2.4. Subsequently, we briefly note organizational matters related to practical implementations of information flows in Section 4.3. In Section 4.4, we then discuss the suitability of proposed technical building blocks based on different supply chain characteristics and the dimensions of information flows. To aid this discussion, we summarize these supply chain characteristics here in Table 2. Based on this overview, we highlight the technological gap when realizing information flows considering given supply chain characteristics in Section 4.5. Thereby, we (1) formalize an abstract and reusable taxonomy for future use, and (2) outline relevant research aspects where successful collaborations between supply chain experts and computer scientists would be highly beneficial.

Table 2. Overview of essential supply chain characteristics (cf. Section 2.2.1) that determine business relationships

| Characteristic | Definition |
| --- | --- |
| Structural Complexity | Determined by the number of actors which are involved in a supply chain and their (indirect) interconnections through business relations |
| Geographic Location | Supply chains range from local to global |
| Duration of Collaboration | Business relations range from short-term to long-term |
| Dynamism | Expresses whether the set of actors is static or dynamic |
| Trust | Outlines the trust relationships between actors |

They greatly influence the circumstances and technological building blocks of information flows.

Table 3. The dimensions *data*, *security*, and *utility* must be considered when talking about reliable information

| | Property | Definition |
| --- | --- | --- |
| Data | Volume | The amount of data that is shared |
| | Velocity | The frequency with which data is shared |
| | Variety | The types of data that are shared |
| Security | Confidentiality | Measures against unauthorized access (privacy) |
| | Integrity | Measures against unauthorized manipulation |
| | Availability | Measures to ensure accessibility and flow of data |
| Utility | Accountability & Verifiability | Information is clearly attributable, and received information can be (independently) corroborated |
| | Authenticity | Information is considered legitimate and genuine |
| | Durability & Timeliness | Information is still valid (accurate + usable), and received information is on time for further use |
| | Liability & Safety | Information use is conformable to public law and steers clear of unwanted dangerous (side) effects |

### 4.1 An Information Flow Taxonomy

Based on our conducted meta-survey, we now derive a taxonomy that focuses on established information flows for the communication and exchange of data within supply chains.

**Methodology.** We build upon the aspects discussed in the different survey articles for our taxonomy. We first collected all referenced properties before merging related properties under a single definition and, finally, grouped the remaining definitions into three distinct dimensions. However, we also reference the original properties throughout our discussion to provide an overview of existing terms. As these discussions are scattered across the literature and emphasized to varying degrees, we provide exemplary pointers here and consider fully embedding these articles into our taxonomy out of scope (cf. Section 3).

We explicitly base our proposed taxonomy on well-known terms and models from the domain of computer science to

(1) ease its understanding in general and
(2) improve the recognition of our taxonomy through its simplicity.

**Overview.** Table 3 provides an overview of our taxonomy. At its core, we identify three relevant dimensions of reliable information flows that align well with the core concepts of their respective subdomains. First, the *data dimension* covers all aspects related to the flow and shape of exchanged information and is characterized by the *3 Vs of Big Data* [153]. Second, the *security dimension* captures relevant criteria of data and information security and can be expressed via the well-known *CIA triad* [200]. Third, in the *utility dimension*, we group properties loosely related to the *quality of the information flows* [139]. These properties are motivated by attributes that describe the data quality in other areas, i.e., we adopt this view for information flows in supply chains. In the following, we present each dimension in more detail.

**Data Dimension.** The first dimension, the *data dimension*, provides an abstract view of the shape of exchanged data to allow for properly expressing corresponding challenges of the information flow in question. Instead of focusing on the monetary value or information communicated via the data, we here rely on the *3 Vs of Big Data* [153], i.e., volume, velocity, and variety, and emphasize the associated processing needs.

Traditionally, related work captures Big Data only for the local processing of information. However, with increased interconnection and information sharing, this dimension is also crucial for established information flows between supply chain actors. In our survey, we further came across a variety of (corresponding) terms, ranging from "growth" [8] over "capacity" and "breadth" [19] to "scalability" [9, 12, 99] when referring to volume (e.g., used by [9, 137]). Contrarily, velocity was discussed less prominently: Apart from the synonym "frequency" [206], we only attribute frequency-related aspects of scalability to this property. As an alternative for variety (e.g., used by [9]), we found the use of "heterogeneity" [175].

**Security Dimension.** Especially with several stakeholders and the flow of sensitive information, security is a crucial challenge. Thus, based on the well-known *CIA triad* [200], we also define a *security dimension* for information flows.

Literature from supply chain experts frequently refers to related concepts only collectively as "security" [12, 99, 115, 175, 206, 210]. Apart from a subset of the three terms known from the CIA triad [77, 170, 182, 193], authors frequently also refer to "(data) privacy" [12, 51, 99, 175, 186, 206, 210] or "data access" [170], which are related to data confidentiality. Then again, "access control" [170] and "authentication" [20, 170] occasionally capture the combination of confidentiality and integrity. Concerning availability (e.g., used by [77, 193])—a rarely explicitly covered topic—we also noticed discussions about "single points of failures" [119], the role of "reliability" [166], or advocating for "decentralization" [199].

However, when developing technical building blocks to secure information flows, a more fine-granular view of security is needed not only in light of privacy concerns but also from a safety perspective. Thus, we capture this "general" aspect using the properties of confidentiality, integrity, and availability. Here, availability should be considered more broadly when compared to the traditional CIA triad. Additionally, the need for delayed information flows and resulting long-term data availability requirements should be considered. Naturally, other best practices for data security, such as data minimalism, are relevant as well when describing and implementing information flows. However, these soft criteria are not directly applicable when categorizing information flows.

**Utility Dimension.** Data quality and utility are often key concerns of information management, and they directly translate to information flows, e.g., to capture details on the origin of data. Consequently, we augment the data and security dimensions with a *utility dimension*. Here, we intend to abstract from content, i.e., the value for the supply chain process itself and focus on "hard" properties of the information flow instead. The surveyed articles discuss the related properties heterogeneously, and hence we group them more broadly but in line with previously proposed attributes [139] as follows.

*Accountability & Verifiability.* This property covers all aspects related to the lifecycle and path of the information flow and its associated data. Overall, any data should be clearly attributable to a party to ensure accountability. Hence, this property also covers the traceability of data across information flows. In terms of auditing, verifying the legitimacy and correctness of data is important, e.g., to discover faulty data. Thus, information flows might have to account for these needs as well. Depending on the use case, ensuring that this property is fulfilled in the long term can be highly beneficial, e.g., when dealing with product faults after decades of usage.

Apart from "accountability" [12, 112] and "verifiability" [12], related work also occasionally discusses the aspects of "non-repudiation" [170], "identification and certification" [19], and "trust" (based on accountability) [116] in this regard.

*Authenticity.* When further considering the origin of information, authenticity is another important property. Companies desire that information is legitimate and genuine. In settings where the origin cannot be properly identified, data tampering cannot be reliably excluded. Consequently, information flows should ensure that authenticity can be verified. We chose the umbrella term based on how the majority of authors refer to this property (e.g., found in [20, 112]). Further, lack of information "transparency" [8], "credibility" [72], or "provenance" [51] are used rarely.

*Durability & Timeliness.* Especially for information flows that rapidly deliver updates, exchanged data might only be valid for a short period, i.e., data becomes obsolete, inaccurate, or unusable over time. Hence, the information flow must allow for frequent updates in such a setting. Furthermore, depending on the use case, low-latency information flows might be required to ensure usability. Thus, we can also characterize information flows according to their latency and timeliness. Consequently, we capture these aspects in a property.

Most frequently, surveys talk about "timeliness" [12, 72, 137, 193] or "timely information" [8, 19] as well. As the most prominent alternative, we discovered the expression "real time" [9, 45, 137]. Finally, we encountered "delivery" (to express speed and dependability) [212].

*Liability & Safety.* The last property for the utility dimension mainly has relevance for legal ramifications that potentially have an implication on the technical solution as well. First, implemented information flows should be conformable to public law. Otherwise, involved companies should be liable for their actions (cf. accountability). Second, when operating and utilizing information flows, they should be safe for the operators, humans, and the environment. For example, an interruption of the information flow should not result in a disaster. Thus, safety also contributes to this property.

This property is rarely featured in related work. Primarily in the context of food supply chains, "safety" (or labeled as "accuracy" [8]) is considered [19, 44]. Furthermore, surveys talk about missing policies, regulations, and legislation [99, 115], which we include under this umbrella as well.

**Implications of our Taxonomy.** To conclude, only when combining these three dimensions, we can uniquely characterize and describe information flows within supply chains. These flows create the foundation to exchange valuable data between stakeholders, where the aspect of information and data quality is key (e.g., [9, 77, 137, 157, 175, 193, 212]). Hence, building upon this base, the aspects of value and veracity become very important for practitioners. However, only when having access to reliable information (flows), the added value from a business perspective can be created and reliably implemented, e.g., through improved decision-making. Thus, our taxonomy with its properties is key when talking about reliable information and evolved, digital SCM.

### 4.2 Application of the Taxonomy to Common Use Cases in Supply Chains

After establishing our taxonomy on information flows, we now look at exemplary applications for the previously outlined use cases (cf. Section 2.4). Even though all dimensions should be considered when implementing a use case, their importance varies depending on the specific use case (and the individual preferences of the involved stakeholders). Accordingly, in the following, we highlight which use cases are particularly affected by the different properties of our taxonomy.

**Data Dimension.** The data dimension and its properties are of particular importance when various actors and information sources are involved. On the one hand, the management of critical infrastructure (UC7) deals with a wide variety of different data from various sources from around the world. Hence, the properties *variety* and *volume* must be given special consideration. On the other hand, real-time monitoring (UC3) is characterized by a high amount of generated data that must be shared between various actors. Moreover, for real-time monitoring, information must be shared quickly and frequently. Thus, this use case commonly has high demands on the *velocity* and *volume* of information flows. When sharing detailed product or production information with customers (UC8), special attention needs to be given to the data's *variety* since vastly different types of information may be shared to accommodate the needs of every supply chain actor.

**Security Dimension.** Even though security is crucial in every use case where information is shared between different actors, some properties are especially important for certain use cases.

Collaborative planning (UC1) and supply chain design (UC2) deal with information that is critical (and sensitive) for the business of multiple stakeholders. Similarly, when sharing product information (UC8), sensitive data on product details and production setups is processed as part of the information flow. All of these use cases handle highly sensitive information, such that its *confidentiality* must be ensured. Further, the *integrity* property is especially relevant when validating entire products or certain characteristics (UC6) as unauthorized manipulations must be prevented. For information flows related to the design and structure of supply chains (i.e., its business relationships and the production composition), the *availability* of information is a primary concern. Especially the sourcing (UC5) and handling (UC4) of faults both require continuous information flows. Thus, for these use cases, the *availability* property is particularly important.

**Utility Dimension.** As described before (cf. Section 4.1), the utility dimension refers to "hard" properties of an information flow. Here, special attention needs to be paid to *accountability & verifiability* of information when relying on accurate product tracing data (UC4–UC6). In addition to the relevance of integrity when validating tracing information (UC6), the *authenticity* of said information flows is equally important. For collaborative planning (UC1), decisions need to be based on current developments, emphasizing the particular importance of the *durability & timeliness* property. Likewise, in the context of critical infrastructures (UC7), information needs to remain actionable. Finally, the *liability & safety* of information flows is also highly relevant for the validation of product tracing data (UC6) as this use case is largely influenced by regulations and contracts.

While all properties from our taxonomy are important for information flows in supply chains, mapping these use cases to our taxonomy highlights the vastly different demands of different use cases. In practice, real-world realizations of information flows are further challenged by stakeholder-specific preferences. Consequently, when integrating information flows into inter-organizational processes, a variety of valid approaches is expected based on the needs of specific supply chains (and their actors). Ideally, proposed and field-tested solutions can be expanded and integrated over time to realize more general and flexible solutions for reliable information flows that satisfactorily realize all properties outlined in our taxonomy, even for the most stringent requirements.

### 4.3 Integrating Information Flows into Inter-Organizational Processes

Several operational aspects become important when integrating information flows into inter-organizational processes. However, they are not primarily relevant for the flows themselves, but rather for the overarching organization and management of the supply chain (network).

**Operational Dimension.** Several properties affect the (local) success, acceptance, and market penetration of recently integrated information flows in practice. They mainly relate to obstacles in effectively sourcing, utilizing, and sharing data.

Apart from interoperability challenges of prevalent information systems [20, 99, 152] and their technical realization, the operational dimension is also affected and inhibited by power relations between different supply chain stakeholders [32, 150]. To circumvent such relations, stakeholders have to be integrated into a joint governance and compliance concept, and thus, they must implement the information flows and information sharing with clearly defined roles, rights, and obligations [49, 51, 76, 138]. Importantly, newly implemented information flows need to establish added values for each involved actor. One common approach to motivate actors to participate, and to realize an added value, is the use of incentive mechanisms [52, 206].

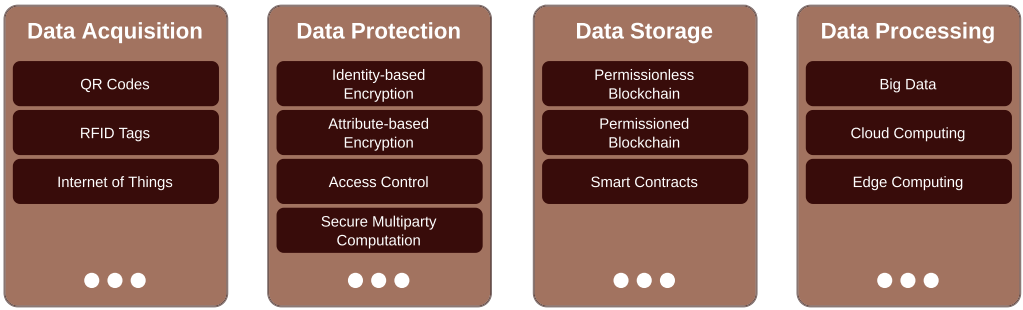

Fig. 6. The data lifecycle encompasses acquisition, protection against unintended access or manipulation, secure long-term storage, and processing to gain insights. For each stage, we identify major technical building blocks that promise to improve the reliability of information flows. However, they have not yet been rigorously implemented in the context of supply chain-specific information flows.

The added value not only follows from the value of shared information but also depends on the costs that occur by establishing the information flows [36, 112, 117, 122, 206]. In addition to initial setup costs, additional costs for the communication infrastructure and the processing of shared information must be considered. Thus, the cost property significantly influences the operational dimensions, and in turn, the establishment of information flows and the exchange of data.

**Takeaway.** Thus, when integrating advanced information flows into (potentially already established) supply chain networks, considering both technical and operational aspects is important. Through our information flow taxonomy and the overview of operational aspects, we provide a foundation and a guideline for assessing and modeling existing supply chains and describing relevant use cases for information flows along with their characteristics. Based on these specifications, suitable building blocks for communication can be identified and finally combined for realizing and implementing thorough, yet tailored technical solutions to realize reliable information flows. With our subsequent discussion on literature-proposed technical building blocks, we now outline their suitability for specific supply chain characteristics and information flow requirements along with relevant implications of their usage.

### 4.4 Proposed (Technical) Solutions in Practice

Our SLR in Section 3 highlights that a multitude of technological building blocks has been suggested to establish information flows in supply chains. However, we also identified a notable distance in how the surveyed articles' authors assess them. In this section, we thus revisit the data lifecycle from acquisition to processing and, for each step, identify and discuss underlying core problems and the applicability of specific technical building blocks. We base our assessment on the essential supply chain characteristics (Table 2) and the identified dimensions of information flows (Table 3). In Figure 6, we provide an overview of the main available building blocks that are suitable for addressing the core problems of information flows in supply chain networks. Our observations further serve as a basis for identifying current gaps in addressing these core problems later on.

**Data Acquisition.** As an initial step, supply chains require solutions linking physical flows to their digital counterparts. In this regard, *QR codes* and *RFID* tags [20, 30, 34, 70, 137, 193] offer reliable identification, i.e., such technologies enable automatic tracking and tracing of objects and thus reduce data acquisition errors. Moreover, they inherently provide basic means of accountability and verifiability. Particularly food supply chains benefit from nearly real-time inventory data and product quality information [20, 34]. However, obtaining such associated data requires active scanning of QR codes and RFID tags at discrete points in time and thus might contradict the desired timeliness of information flows [70].

In turn, the **Internet of Things** *(IoT)* extends the idea of linking the physical to the digital world by introducing sensors that *actively* report measurements using wireless links and thus fully automate said process. For supply chains, IoT-based solutions may cover high volumes, velocities, and varieties of data and offer means to provide confidentiality, integrity, and availability. Nevertheless, the adoption of IoT technology is generally slow due to high initial costs and high perceived complexity [45, 58, 67, 115, 119, 166, 175]. Moreover, battery-powered IoT devices might struggle to satisfy desirable security needs in practice.

**Data Protection.** After the acquisition, the data has to be prepared for storage and further utilization. With additional (shorter-lived) collaborations, i.e., a high level of dynamism, come elevated risks of accidental and direct or indirect data and knowledge leaks [210], for example, untrusted parties gaining illegitimate access, or (previous) access rights are not fully revoked (on time). These associated risks of losing control over one's data or the need to deliberately give up this control primarily hinder large-scale information exchanges [12, 70, 170]. Available cryptographic building blocks, such as encryption (including identity- or attribute-based encryption [12]) or access control, promise to mitigate unintentional data leaks. However, collaboration may require collaborators to share their knowledge to create further insights. In cases where the collaborators do not fully trust each other, i.e., perceived risks primarily originate within the supply chain [32], *SMC* can be applied to help collaborators compute the result of a function based on confidential inputs without having to reveal these inputs [210]. Unfortunately, the performance of SMC-based protocols decreases quickly with increasing complexity or inputs [69], i.e., the application of SMC in its current state for large supply-chain settings currently requires intensive analyses of feasibility, costs, and benefits.

**Data Storage.** Data exchanges require that the data is available to all intended receivers. In recent years, we have observed an increased interest in the application of *blockchain technology* and *smart contracts* to facilitate the coordination between collaborators in (especially complex and globally distributed) supply chains [12]. Blockchain-based solutions are especially promising in settings where trust between collaborators has not been fully established yet, e.g., in settings where supply chains are highly dynamic. Blockchain systems can be either *permissionless* or *permissioned*, i.e., publicly available to anyone or accessible by a mutually known set of collaborators. Even though permissionless blockchains allow for much more flexibility in dynamic settings, while providing public verifiability of what data has been recorded in the past, these systems have to resort to resource-intensive consensus protocols to cope with the higher risk of interference by untrusted participants. Contrarily, permissioned blockchains allow for more fine-grained and more efficient control over the exchanged information, e.g., mutually known collaborators can establish data authenticity more easily. However, this approach does not allow for the same dynamism as permissionless blockchains do. In conclusion, the means to operate in a decentralized manner with (only) partly trusted collaborators highly depend on the parameters of the underlying supply chain and the required flexibility regarding potential collaborators.

So far, blockchains are predominantly deployed to improve traceability in food supply chains [97]. However, other sectors (e.g., textile industry) have started to explore their potential [119]. Especially food safety is an often-cited driver for the use of blockchain-based information exchange [97, 119]. As the deployment of blockchains to support supply chains increases, the volume, velocity, and variety of blockchain-recorded data are bound to increase. However, unresolved scalability concerns of blockchain frameworks, as well as a lacking standardization in this field [116], impose further challenges for the widespread adoption of blockchain-backed supply chains, especially if the supply chains experience high levels of dynamism. This lack has also been identified for general supply-chain solutions [20].

Smart contracts can serve as an interface to codify and thus automate contractual rules, e.g., conditional payments, and thereby alleviate the complexity of blockchain-backed collaborations. However, growing validity and security concerns challenge their use [116], and they publish codified agreements to all partners. If data confidentiality among the collaborators is a priority, the discussed building blocks for data protection are needed to ensure the privacy of information flows as well.

**Data Processing.** Finally, the data will be processed to gain further utilizable insights. With structurally complex supply chain networks and high data volumes, special requirements for technical solutions arise that should handle complex and extensive data. With increasing volume, velocity, or variety (cf. Table 3), *big data*-based solutions are discussed frequently [32, 99, 171, 175, 213]. However, given the widespread confidentiality and availability concerns, a distributed processing of data might be more suitable.

To realize data processing, related work repeatedly mentions two basic concepts. *Cloud computing* has been identified as a promising building block when dealing with a complex data dimension [213] due to its flexibility [30] and scalability [9], but also for the ease of extending the set of stakeholders. However, as a (logically) centralized concept, cloud computing introduces a single point of failure, where cyber-attacks or other outages can have an enormous negative impact on the security and utility dimensions [83]. As an alternative, *edge computing* [44] follows the principle of decentralized data processing with a centralized purpose while still focusing on added value for all involved stakeholders, i.e., the processing is closer to the origin or the recipients of specific information, which increases capabilities regarding data volumes while further dealing with confidentiality concerns. However, new technology, e.g., edge-based setups, might increase coordination overheads and might even deter companies from participating [70, 170]. Regardless of the approach, standardized formats [30], approaches [82], and interfaces are needed to realize inter-organizational collaborations [115]. Still, more work in this direction is needed [51].

In conclusion, computer scientists have developed a multitude of technical building blocks that seem suitable to also improve the reliability and confidentiality of data exchanges in supply chains at every stage. However, the complexity of modern supply chains and their highly individual requirements necessitate further tailoring of these building blocks in this challenging scenario. In the following, we thus take a closer look at this technological gap.

### 4.5 Technological Gap

The technical solutions that we compiled with our meta-survey (cf. Section 3) address various aspects and challenges for information flows within supply chains. Despite their individual strengths and potentials, open challenges remain for various combinations of supply chain characteristics and information flow properties. To the best of our understanding, these challenges primarily arise from technological gaps, which in turn follow from both missing technical building blocks as well as a lack of shared inter-domain knowledge.

**Supply Chain Complexity.** An important finding is that today's solutions were initially and mainly proposed for use in small, static supply chains. Thus, they fail to satisfy the requirements of modern, complex, and rapidly evolving supply chain networks [20]. Especially, the security and utility dimensions must be carefully revisited in light of multi-hop collaborations, i.e., a transformation from local collaboration clusters to global networks [107] is not only needed from a business perspective, but also from a technological one.

**Reliability-by-Design.** As in other settings, security and privacy are unfortunately still mostly regarded as an unwanted necessity, i.e., when selecting technologies for deployment and means of communication, the concepts of security-, safety-, and privacy-by-design are rarely considered. Further, concepts that target the authenticity of submitted data [128] still lack practical implementations. Finally, technical solutions for reliable retraction of shared information, e.g., at the end of an extensive collaboration phase, are missing. Thus, proposed and developed building blocks as well as thorough approaches frequently neglect these aspects in their entirety as well. Consequently, today's solutions frequently fail to consider all relevant information flow properties, potentially raising the bar for their practical deployment. However, without a doubt, depending on the use case, their importance varies.

**Inter-Domain Collaboration.** Altogether, we notice that computer scientists cannot tackle all raised aspects on their own (cf. organizational dimension). Still, upcoming research efforts should pursue advances in supply chains more holistically, primarily by considering both the supply chain characteristics and the different properties of our taxonomy when developing, evolving, and proposing (new) building blocks, methods, and models for use in supply chains. Moving toward a more abstract view of general challenges at the intersection of supply chain and computer science research, we next discuss the most crucial direction in Section 5.

## 5 FUTURE RESEARCH DIRECTIONS

Based on the insights generated from our literature meta-survey (Section 3) and our own research experience in the area, we now discuss and motivate relevant future research directions toward more secure and reliable information flows along supply chains. With this background in mind, our discussion focuses on necessary steps to close the inter-domain gap between industry and scientific perspectives, especially computer science, to promote comprehensive and sustainable solutions.

Already in 2010, Sarac et al. [156] criticized the separation of supply chain literature into industry-specific and academic works. Due to the lack of details on corresponding technical solutions in the surveyed articles, we specifically complement our findings with directions identified by additional related work to achieve a thorough and well-founded collection of research directions. Even though several of these aspects are already well-addressed within the computer science research community on a conceptual or isolated level, use-case-specific requirements and challenges result in novel, still open research areas when adopting existing technologies and concepts. Thus, we identify a prevalent inter-domain gap impeding the adaption of existing (sophisticated) IT-based solutions. Hence, this section covers both (1) novel computer science research directions and (2) well-known research areas with a need for inter-domain adaptation and refinements. Indisputably, past efforts also proposed valuable approaches at the intersection of computer science and supply chain research. However, to the best of our knowledge, these efforts primarily focus on higher layers (cf. Figure 3(a)), disregarding the (technical) foundations of information flows—primarily concerning the layer *sensing and processing*. That is, they build on (and also require) reliable information flows to truly realize their presented contributions without outlining how to implement them in practice.

In Table 4, we provide a high-level overview of the different research directions and the required inter-domain collaborations along with the respective required expertise (involvement) and preliminary work of both domains. We stylistically distinguish the different properties and rate each property on a scale from one (little) to three (significant) to give an overview of our

Table 4. Despite existing and ongoing, yet primarily isolated, preliminary work in both domains, we identify the significant need for inter-domain collaborations to truly advance the research intersection of reliable information flows in supply chains

| (Future) Research Direction | Req. Inter-Domain Collaborations | Computer Scientists | | Supply Chain Experts | |
|---|---|---|---|---|---|
| | | Involvement | Preliminary Work | Involvement | Preliminary Work |
| Refining Information Flows | 💬💬💬 | ☐☐ | ⊘⊘⊘ | ☐☐☐ | ⊘⊘ |
| Implications of Governance | 💬💬 | ☐☐ | ⊘ | ☐☐ | ⊘⊘⊘ |
| Mitigating Opportunistic Behavior | 💬💬 | ☐ | ⊘⊘ | ☐☐ | ⊘⊘ |
| Transparency vs. Confidentiality | 💬 | ☐☐ | ⊘⊘⊘ | ☐ | ⊘ |
| Improving Data Reliability | 💬 | ☐☐☐ | ⊘ | ☐ | ⊘ |
| New Technologies for SCM | | | | | |
| *Adoption potentials* | 💬💬💬 | ☐☐ | ⊘⊘⊘ | ☐☐☐ | ⊘⊘ |
| *Standardization: SCM-as-a-Service* | 💬💬💬 | ☐☐☐ | ⊘ | ☐☐ | ⊘ |
| *Findability of Data and Information* | 💬 | ☐☐ | ⊘⊘ | ☐ | ⊘ |
| Sophisticated Evaluations | | | | | |
| *Universal Supply Chain Models* | 💬💬💬 | ☐ | ⊘⊘ | ☐☐☐ | ⊘⊘ |
| *Availability of Real-World Testbeds* | 💬💬 | ☐☐ | ⊘ | ☐☐ | ⊘ |
| *Longitudinal Studies* | 💬💬 | ☐☐☐ | ⊘ | ☐☐☐ | ⊘⊘ |

☐: little  ☐☐: moderate  ☐☐☐: significant.
The availability of preliminary work in one domain does not necessarily entail that less future involvement is needed.

impression w.r.t. specific research directions: While we expect the need for significant contributions by computer scientists in some directions (e.g., improving data reliability), other directions (e.g., the development of supply chain models) have a stronger focus on the supply chain background. Overall, our table underlines the significant need for interdisciplinary collaborations, where the required (research) input depends on the concrete direction. These insights motivate our following discussion.

**Refining Information Flows.** The deficiencies regarding inter-domain concepts for information flows identified by our meta-survey call for an interdisciplinary treatment of these concepts, covering deployment and management aspects of supply chains. First, these deficiencies must be addressed by formalizing the concepts of information flows to allow the seamless cooperation of interdisciplinary teams on their application to realistic problems. Then, the technical challenges of integrating these developed concepts into deployable systems must be addressed. Thus, future research on supply chain information flow should address the following two questions.

▶ *How can the different perspectives and requirements of information flows be formalized in an interdisciplinary fashion?* While research on information flow concepts is not novel per se, the required novelty stems from exchanging domain-specific knowledge and requirements with a common inter-domain taxonomy (cf. Section 4) to initially derive, extend, and ultimately formalize feasible information flow concepts. Here, a more methodical and formal treatment of the present requirements and restrictions from the supply chain domain by all stakeholders paves the way for computer scientists to propose thorough, feasible, and viable solutions for sophisticated information flows and business processes within and along supply chains.

▶ *How can these refined information flows be realized in practice?* Today's technologies can result in improved decision-making when being combined with emerging technologies, new algorithms, and specific use cases [30, 99, 213]. Currently, such opportunities are, however, often under-utilized, as their full potential can only be unleashed through inter-domain collaboration. While a formal treatment of information flow concepts can identify where novel information flows can bring disruptive changes, deploying these advances poses at least as big of a challenge. Not only systems realizing these flows handle enterprise-sized processes, but they must also be deployable seamlessly and most likely incrementally into functioning supply chains with many stakeholders having different deployment processes and requirements, e.g., regarding privacy [194].

**Implications of Governance.** Despite the potential (theoretic) benefits of evolved supply chains, considerations regarding governance and standardization in practice constitute a highly relevant topic. While computer science-driven solutions often are decentralized and designed without a controlling third party, companies frequently demand central monitoring capabilities (partly due to regulatory purposes). For example, compliance verification is mandatory (during tracing and tracking) and might assist at the time of recalls [180]. Hence, future work should look into governance frameworks that allow for a certain degree of autonomy, regulation, and legal penalties yet combined with information confidentiality (cf. security dimension, Section 4.1). In particular, sophisticated or even optimal solutions from a computer science perspective require changes that respect governance and legal requirements. Consequently, technical measures that provide regulators with reliable access to the required information while prohibiting them from analyzing the content should be developed while also considering the following questions.

▶ *How can supply chain actors motivate their partners to increase the visibility of their information flows for regulatory purposes?* To date, privacy concerns challenge the establishment of additional information flows. Thus, future research should address these aspects to ultimately push for real-world use, e.g., using incentive mechanisms that reward actors who share crucial information [104].

▶ *What is the optimal tradeoff between human and technological resources for successfully implementing reliable information?* Conceptually, technology can only support governance. However, governance cannot be automated entirely [41, 126]. Consequently, techniques and strategies are required to effectively analyze the human in the loop to govern, provide regulation, and make decisions wherever necessary.

**Mitigating Opportunistic Behavior.** At times, supply chain actors only invest in traceability systems to mitigate their internal risks and do not consider implications on external partners [176], seeking a local financial optimization with potential global long-term shortcomings [168]. Thus, more emphasis should be put on models that represent such changes holistically, i.e., minimizing the (global) implications of opportunistic behavior. Existing research has identified game theory as a promising technique to analyze and influence the opportunistic behavior of individual stakeholders [207]. That said, future research should further try to derive universal incentive models that balance internal and external benefits.

**Transparency vs. Confidentiality.** Distantly related to opportunistic behavior, information transparency and confidentiality directly oppose each other and significantly influence how businesses benefit from sophisticated information flows and corresponding business relationships [147] (cf. the mentioned concerns in Section 3.5). In practice, corresponding information flows and pressing concerns may still differ across supply chains due to different views of confidentiality as well as cultural and legal aspects [124]. Thus, these views may also affect the level of openness [90] and transparency, such that further research needs to consider them when developing inter-domain solutions that are guided by the following questions.

▶ *Which risks are associated with sharing information?* Proper risk assessments on sharing information with collaborators are needed to establish information flows in practice. Existing literature either focuses on the confidentiality or the availability of the data and very few (e.g., [104, 106]) discuss both aspects while maintaining the utility of the data. In line with this question, researchers should investigate what information must be exchanged at the bare minimum to implement (novel) use cases. Data minimalism is well-known in this context [54] and is frequently applied.

▶ *Which information must be shared to realize minimalistic, and thus efficient and privacy-preserving, dataflows for specific supply chains?* Future research should explore the potential benefits of creating data-sharing standards for supply chain collaborators that focus on accurately and adequately sharing only relevant information. Besides the performance benefits, avoiding over-sharing data also mitigates the severity of privacy issues. Technical solutions to address these issues while maintaining the gained efficiency are needed as well.

**Improving Data Reliability.** To strengthen businesses' advantages of—and hence their incentives for—providing sophisticated transparency, the respectively shared supply chain data, irrespective of its content, must be authentic and trustworthy to allow for reliable decision-making in practice. Suppose the available data is of low quality, faulty, or tampered with in any way. In that case, improper or costly business decisions likely arise, potentially raising justified concerns against data sharing. Hence, future research must address the following questions to overcome the risks resulting from unreliable data sharing and utilization (along supply chains).

▶ *How can we ensure the reliability of supply chain data?* When looking at the complete data lifecycle, measures and concepts are needed to handle the data with care at all times, which might eventually improve reliability, trust, and data quality as well. Initially, the data acquisition and sensing must be secured, e.g., through trusted sensing [128], especially when handling goods in untrusted environments. **Trusted platform modules (TPMs)** and **trusted execution environments (TEEs)** can help to aid secure data exchanges and processing [11, 75]. So far, corresponding approaches are still in their infancy. Data reliability can also be improved by adopting strong cryptographic measures, such as data consistency through hash functions in blockchains [105, 205]. Similarly, trust models [39, 105] known from information sharing can also be adopted for reliable communication [50].

▶ *How can we reliably map product and information flows?* In light of data reliability needs, we look forward to further evolution of initial approaches for tamperproof markings of shipments and products (e.g., molecular fingerprinting, smart fingerprints, laser markings, and others) [140, 173, 191, 211, 215]. Such smart fingerprints (e.g., [88]), for example, cannot be counterfeited and thus can be very useful in conjunction with trust models, trust architectures, and modern digital technology, for validation purposes and for identifying faults (cf. Section 2.4). Third-party certificates attesting the product quality can also improve reliability [6]. They can also be digitized for (automated) usage in information flows. The first examples cover pharmaceuticals, diamonds, organic produce, and wherever ethical and sustainable practices [146] in product manufacturing demand verification.

**New Technologies for SCM.** The operation of existing and future technical solutions for reliable information flows within supply chains needs to be coordinated to allow for smooth operation and for gradual adoption. Consequently, corresponding influences need to be well-researched.

*Adoption Potential.* Newly proposed approaches have to be efficiently and effectively deployable in real-world supply chains to introduce benefits. First, organizations must establish their processes for long-term operations to attenuate spent investment costs over time. Second, introducing new technologies (cf. Section 4.4) constitutes further investments and bears the risk of disrupting any of the carefully established and validated supply chain processes. Hence, organizations tend to be reluctant to explore the potential benefits of emerging new technologies in practice.

▶ *What are the organizational barriers when adopting new technologies, both cost-wise and beyond costs? How can we overcome them?* Detailed analyses of the associated direct and indirect costs must be conducted when adopting new technologies that enable new use cases, for example, through information flows [29, 70, 122, 206]. Thus, they should also cover the aspects of decision-making and training of the workforce when deploying a new technology [9]. In addition, organizations need to factor in the potential revenue as well as long-term benefits following the adoption of new technologies. These cross-domain challenges can only be tackled with appropriate technical expertise (cf. our taxonomy).

▶ *To what extent can digital product verification be improved using new technologies?* Involved organizations and businesses also need to factor in advances in data reliability following the adoption of new technologies, i.e., they need to confirm that they can still satisfy the required certification and data integrity. Thus, we call for additional research to compare the effectiveness of newly proposed technologies. Today's mix of varying maturity, at different operational levels, and various scales of adoption over the years, require a more standardized assessment of their potential, primarily focusing on the presented dimensions of data, security, and utility [147].

*Standardization: The Road toward SCM-as-a-Service.* This contemplated integration of new technologies requires establishing processes that facilitate both the technologies' adoption and the onboarding of affected businesses. Corresponding reliability improvements of information flows, and more generally, computer science-induced advances in the information lifecycle, promise the potential to develop SCM-as-a-Service-like approaches that are based on digital, reliable, and accountable SCM. Thus, traditional supply chain actors could be encouraged to outsource parts of their current SCM, e.g., data analytics, storage, blockchain nodes, or ML/AI-based decision-making when provided with a sophisticated yet comprehensible standardized information exchange and processing infrastructure. When researching this direction, future work can greatly benefit from foundational work that promotes and fosters interoperability in supply chain networks, in parts through interdisciplinary efforts. In addition, where applicable, cross-domain standards [215] can also be adopted more widely to make information flows generally more interoperable [64]. However, to ensure that these advances succeed in practice, research must holistically solve the distinct technical challenges of all layers (cf. Figure 3(a)), including the often neglected lowest layer on sensing and processing of information flows.

*Findability of Data and Information.* When implementing (such) new data-driven processes to improve decision-making (cf. Figure 3(c)), all required input data must be accessible. Due to the increasingly decentralized sourcing of information, companies must ensure that they can find, access, and retrieve all relevant data. Appropriate technical solutions must be deployed in the field, especially in settings where information needs to be shared over multiple hops. So far, research still prioritizes the local view at times [80], and thus, corresponding solutions are not widely applicable.

**Sophisticated Evaluations.** Given the emergence of concepts and technical solutions for realizing reliable information flows, evaluating their costs, applicability, performance, data reliability, and security features constitute another crucial inter-domain research direction.

*Universal Supply Chain Models.* As a first important measure, having access to standardized supply chain models would be highly beneficial for both the development of novel solutions as well as their evaluation, especially when comparing different approaches. However, so far, no such model exists, hindering real-world feasibility studies as well as a standardized way to compare new approaches to existing work. Such information is essential to fully tap into the envisioned benefits (and to deploy research concepts and prototypes into productive use). Since such evaluation models could ease the adoption decisions for companies [156], respective research efforts and proposals are highly encouraged.

*Availability of Real-World Testbeds.* Moreover, standardized supply chain environments that allow researchers to evaluate novel approaches for varying supply chain models are missing. Although existing work already addresses physical testbeds in the context of supply chains and logistics [63], more sophisticated and information flow-focused approaches are yet to be developed. Overall, this shortcoming concerns both reliable sensing, i.e., to evaluate reliable information processing, and testbeds to study the effects of revisited decision-making (e.g., a setup to source realistic information from). Integrating new technologies (including different sensing devices, blockchain technology, or storage solutions) and processes, which do not exist across existing supply chain networks, thus poses a significant barrier, especially in less technology-experienced industry sectors and supply chains.

***Longitudinal Studies.*** While simulations are still the predominant means to evaluate information flows in supply chains [67], empirical research and real-world evaluations are lacking [44]. Most new approaches do not investigate the deployment in the field and over time thoroughly or at all. Therefore, we require more work (and corresponding academic incentives to do so) that also studies the impact of novel solutions in the wild (being positive or negative) [166]. Such a shift would allow researchers and practitioners to better judge approaches regarding their real-world potentials, consequences, and adoption chances. Currently, a barrier between academia and industry seems to exist where industry acts far more pragmatically than 'sophisticated' academic research. Thus, as a third measure toward sophisticated evaluations, we identify improved and deepened collaborations, acceptance, and attribution between those two worlds as highly desirable.

As outlined in this section and summarized in Table 4, the need for future research, particularly inter-domain research efforts, is significant. With our interdisciplinary contributions, i.e., a newly proposed taxonomy and the list of supply chain characteristics, we hope to support upcoming collaborations by providing them with standardized terms and properties. Apart from easing the bootstrapping of collaborations, we further expect that our holistic view of the research area will reduce the risk that researchers accidentally overlook or purposely ignore specific dimensions when researching communication infrastructures and information flows in supply chains.

## 6 CONCLUSION AND OUTLOOK

Increasingly complex supply chain networks imply the need to exchange information reliably. Recent disruptions, such as COVID-19, the Suez canal obstruction, or the Russian invasion of Ukraine, further underline this situation. Only with extensive communication and reliable information can companies make well-informed decisions to deal with disruptions and strengthen their resilience. Surprisingly, as a result of our meta-survey, research in computer science has widely overlooked this aspect. Using analogies from the domain of computer science, we intend to familiarize computer scientists with research in supply chains to eventually foster collaborations with supply chain experts and contribute toward more secure and reliable information flow implementations. Although this article offers only an initial building block for this ambitious goal, it provides a unique perspective on tackling the imminent challenges of supply chains with the help of computer science.

In particular, given the various use cases in supply chain management and their individual requirements, a single technical solution (as frequently advocated) is not a realistic option. Instead, practitioners must compile a precise overview of the needed information for a particular use case and the requirements concerning the corresponding information flow as well as its underlying communication infrastructure. However, to date, a common information flow terminology is missing, which leads to a situation where research challenges are not properly communicated to computer scientists or already existing building blocks are not applied (correctly or at all) in practice. To mitigate this cumbersome situation, we derived an abstract taxonomy, based on an extensive survey, that captures information flows within supply chains using the dimensions of data, security, and utility. Thus, we create a foundation to establish a common understanding of information flows by appropriately referring to research challenges, needs, and strategies.

For future work, we call for a two-fold research agenda. First, appropriate technical building blocks from computer science are required to fully address the challenging needs in supply chains. Second, more general research guidelines (including standardized evaluation models) should be established to improve the comparability of proposed approaches. Thereby, their readiness and potential can be judged more accurately. To conclude, computer scientists can be a powerful driver in advancing information exchange and decision-making in supply chains, with improvements for participating companies and society in general.